\documentclass[fleqn,12pt,twoside]{article}
\usepackage{espcrc1}

\usepackage{graphicx}
\usepackage{amssymb}
\usepackage[figuresright]{rotating}

\newcommand{\AmS}{{\protect\the\textfont2
  A\kern-.1667em\lower.5ex\hbox{M}\kern-.125emS}}
\def\araa{ARA\&A}%
\def\apj{ApJ}%
\def\aap{A\&A}%
\def\aaps{A\&AS}%
\def\pasp{PASP}%
\title{Rotating massive stars: Pre--SN models and stellar yields at
solar metallicity}

\author{
R. Hirschi\address[obsge]{Observatoire de Gen\`eve, 1290
Sauverny, Switzerland},
G. Meynet\addressmark[obsge] and 
A. Maeder\addressmark[obsge]
}
       
\begin{document}

\maketitle

\begin{abstract}
We present a new set of stellar yields obtained from 
rotating stellar models at solar metallicity covering the 
massive star range (9--120 $M_{\odot}$). The stellar models were
calculated with
 the latest version of the Geneva stellar evolution code
described in \cite{psn04a}. 
Evolution and nucleosynthesis are in general followed up to Silicon
burning.
The contributions from stellar winds and
from supernova explosions to the stellar yields were calculated separately. 
The two contributions were then added to compute the total stellar 
yields \cite{ywr04b}.

The effects of rotation on pre--supernova models are significant 
between
15 and 30 $M_{\odot}$. 
Above 20 $M_{\odot}$, rotation may change the radius or colour of the 
supernova progenitors (blue instead of red supergiant) and
the supernova type (IIb or Ib instead of II).
Rotation increases the $\alpha$ and CO core sizes by a factor $\sim 1.5$. 
\textbf{Thus, rotation increases the  
yields for heavy elements and in particular for carbon and oxygen 
by a factor 1.5--2.5}.
Rotating models produce larger yields for $^{12}$C and 
$^{16}$O in 
the mass range between 
9 and about 35 $M_{\odot}$ compared to the 1992 calculations \cite{AM92}.

For Wolf-Rayet
stars ($M \gtrsim 30 M_{\odot}$), 
the pre--supernova structures are mostly affected by the intensities
of the stellar winds and less by rotation \cite{ROTX}. 
In this mass range, 
rotation increases the yields of helium and
other hydrogen burning products but does not significantly affect the yields of
elements produced in more advanced evolutionary stages.
Note that the final masses of the most massive stellar models 
($\sim 120\,M_{\odot}$) are similar to the final masses of less massive
stars ($\sim 40\,M_{\odot}$) due to the use of 
revised mass loss rates from Nugis and Lamers 2000
\cite{NuLa00}. The most massive stars are therefore also expected to 
form black holes. 
\end{abstract}

\section{Introduction}
Over the last ten years, the development of the Geneva stellar 
evolution code has allowed
the study of the evolution of rotating stars until carbon burning. The
models can reproduce many observational features at various 
metallicities, like
surface enrichments \cite{MM02n}, ratios between red and blue 
supergiants \cite{ROTVII}
and the  population of Wolf--Rayet (WR hereinafter) stars 
\cite{ROTX}. In \cite{psn04a}, we describe 
the recent modifications made to the Geneva code
and 
the evolution of our rotating models until silicon burning.
 In this contribution, we briefly present the
stellar yields 
for rotating stars at solar metallicity with
 a large initial mass range (9--120 $M_{\odot}$). 
\section{Computer model}
The computer model used to calculate the stellar models
 is described in detail in \cite{psn04a}.
Convective stability is determined by the 
Schwarzschild criterion. 
Convection is treated as a diffusive process from oxygen burning
onwards.
The overshooting parameter is 0.1 H$_{\rm{P}}$ 
for H and He--burning cores 
and 0 otherwise. On top of the meridional circulation 
and secular shear,
an additional instability induced by rotation, dynamical shear, 
was introduced in the model. 
The reaction rates are taken from the
NACRE \cite{NACRE} compilation for the experimental rates
and from the NACRE website (http://pntpm.ulb.ac.be/nacre.htm) for the
theoretical ones. 
The mass loss rates used are described in \cite{ROTX}.
In particular, during the Wolf--Rayet phase, we use
the mass loss rates by Nugis and Lamers 2000 \cite{NuLa00}. 
These mass loss rates,
which account for clumping effects in the winds,  
are smaller by a factor 2--3 than the mass loss rates used in our 
previous non--rotating, ``enhanced mass loss rate'' stellar grids
\cite{MM94}. 

We calculated stellar models with initial masses of 9, 12, 15, 20, 
25, 40, 60, 85 and 120 $M_\odot$
 at solar metallicity, with initial rotation velocities 
of 0 and 300 km\,s$^{-1}$. The value of 300 km\,s$^{-1}$
corresponds to an average velocity of about 220\,km\,s$^{-1}$ 
on the Main
Sequence (MS) which is
very close to the observed average value \cite{FU82}. 
The calculations start at the ZAMS.
The rotating 15, 20, 25, 40 and 60 $M_\odot$ models were computed until
the end of core silicon (Si) burning. Their non--rotating counterparts
were computed until the end of shell Si--burning. 
For the rotating 12 $M_\odot$ star, the model 
ends after oxygen (O) burning. 
For the non--rotating 12 $M_\odot$ star, neon (Ne) burning starts at
a fraction of a solar mass away from the centre but does not reach the 
centre and the 
calculations stop there. 
The evolution of the models with initial
masses between 12 and 60 $M_\odot$ is described in \cite{psn04a}.
The 9, 85 and 120 $M_\odot$ models are presented in \cite{ROTX} and 
their evolution was followed until the end of the core He--burning
(the SN yields calculation for these last models is described in \cite{ywr04b}
and follows the method used in \cite{AM92}.
\section{Results}
\begin{figure}
\includegraphics[width=8.2cm]{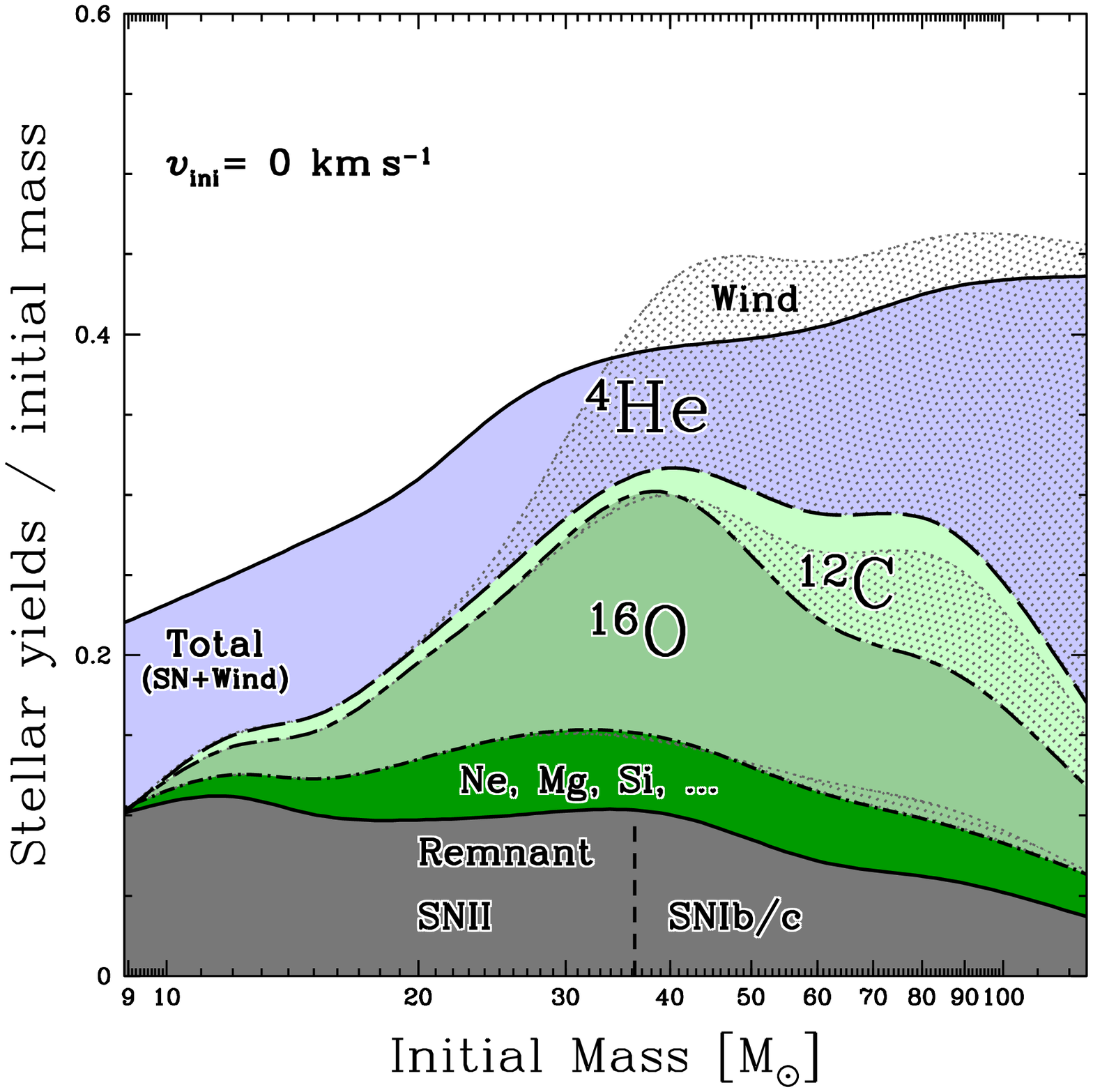}\includegraphics[width=8.2cm]{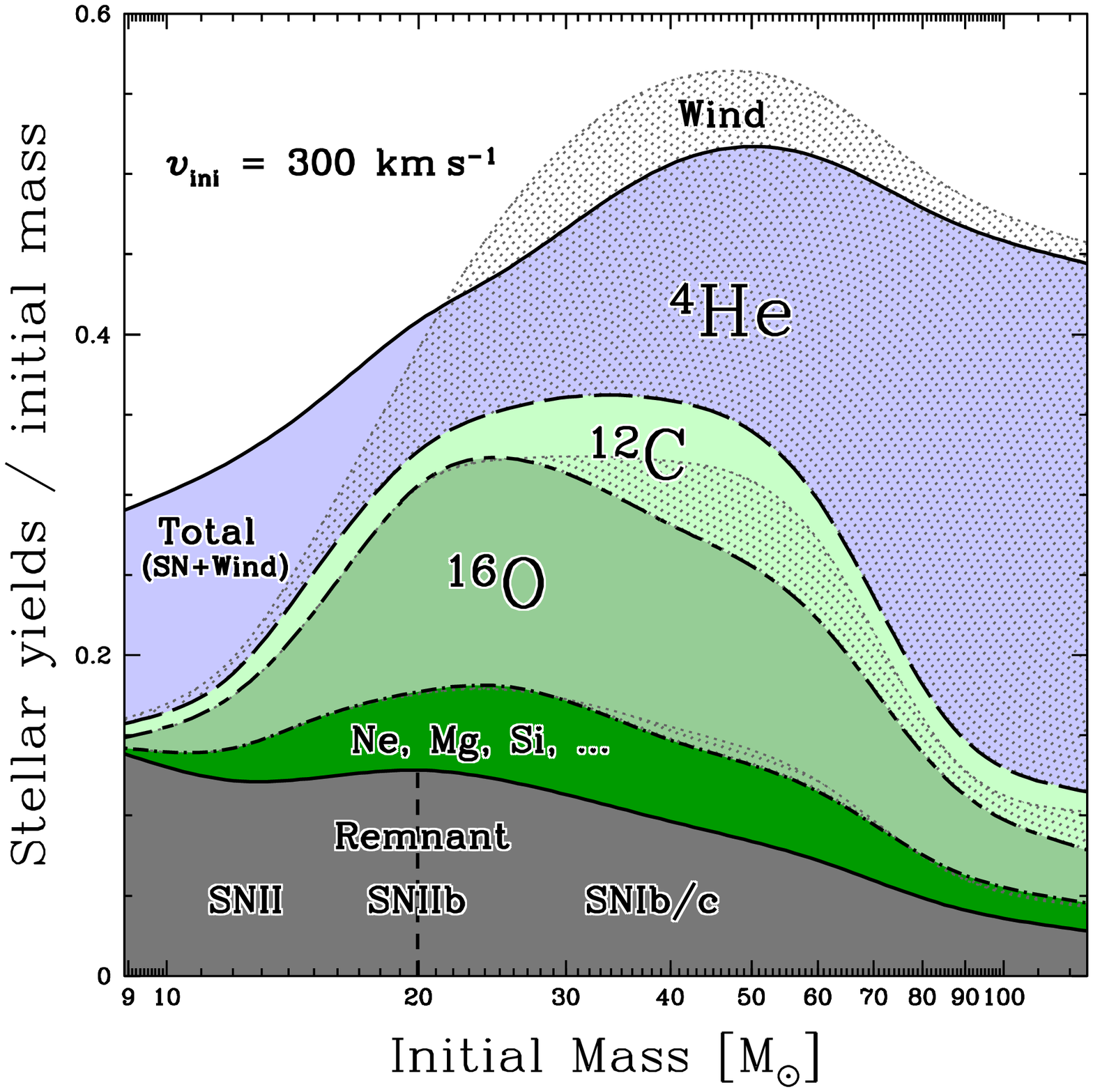}
\caption{Stellar yields divided by the initial mass 
 as a function of the initial mass
for the non--rotating 
 (left) and rotating  (right) 
models at solar metallicity. 
}
\label{ytot}
\end{figure}
\subsection{Contributions to yields from stellar winds and SN explosions}
Before we discuss the stellar yields, it is
useful to recall the influence of rotation on the final mass of 
the different models (presented in
\cite{ROTX,ywr04b}). Below 30 $M_{\odot}$, rotating models lose significantly
more mass than non--rotating models \cite{MM00}.
For WR stars ($M \gtrsim 30 \,M_{\odot}$), the new mass loss 
prescription 
\cite{NuLa00}, including the
effects of clumping in the winds, 
results in mass loss rates that are a factor of two to
three smaller than the rates from \cite{La89}. 
As a result, the final mass of WR
stars in the present calculation are noticeably larger 
than in 1992 \cite{AM92}. There is no clear difference between the final
mass of rotating and non--rotating models.
For a model with an initial mass larger than 30 $M_{\odot}$,  
the final mass is 
always between 11 and 17 $M_{\odot}$. 
Black hole formation is therefore expected for all the very massive stars
at solar metallicity. 

What is the relative importance of the wind and SN contributions? 
Figure \ref{ytot} displays the total stellar yields 
divided by the initial mass of the star 
as a function of its initial mass, $m$, for the non--rotating (left)
and rotating (right) models. 
The different total yields (divided by $m$) are piled up. $^4$He
yields are delimited by the top solid and long dashed lines (top
shaded area),
$^{12}$C yields by the long dashed and short--long dashed lines, 
$^{16}$O  yields by the short--long dashed and dotted--dashed lines and 
the rest of metals by the dotted--dashed and bottom solid lines. The
bottom solid line also represents the fraction of the star locked in the
remnant ($M_{\rm{rem}}/m$). 
The corresponding SN explosion type is also given.
The wind contributions  are superimposed on the total yields for the
same elements between their bottom limit and the dotted line above
it. Dotted areas help quantify the fraction of the yields
due to winds. Note that for
$^{4}$He, the total yields are smaller than the wind yields due to
negative SN yields for $^{4}$He.

For $^4$He (and other H--burning products like $^{14}$N), the wind
contribution increases with mass and dominates for $M \gtrsim 22
M_{\odot}$  for rotating stars and for $M \gtrsim 35
M_{\odot}$  for non--rotating stars. These mass limits correspond to the
lower mass limits for WR star formation.
For very massive stars, the SN contribution for $^4$He is negative (this
is possible because, in
the yield calculation, the initial composition is deducted from the
final one)
and this is why the wind contribution is higher than the total one.
For $^{12}$C, the wind contributions only start to be significant above
the mass limits for WR star formation 
(22 and 35 $M_{\odot}$ for rotating and
non--rotating models respectively). 
This is expected because a star must have ejected most
of its helium before it can eject carbon. Above these mass limits, the
contribution from the wind and the SN are of similar importance.
For $^{16}$O, the wind contribution remains very small because with the
new mass loss prescription, the oxygen rich layers are not uncovered. 

\subsection{Total stellar yields}
Our non--rotating models were compared to the literature 
\cite{RHHW02,LC03,TNH96} and are consistent
with other calculations. Differences can be understood in the light
of the treatment of convection and the rate used for
$^{12}$C$(\alpha,\gamma)^{16}$O \cite{ywr04b}. This verifies the accuracy of our
calculations and gives a safe basis for studying the effects of rotation
on the yields.

For H--burning products, the yields of the rotating models are usually 
higher than those of
non--rotating models. This is due to larger cores and larger mass loss.
However, between about 15 and 25 $M_{\odot}$, the rotating yields
are lower. This is due to the fact that the winds do not expel much
H--burning products yet, and more of these products are burned
later in the pre--supernova evolution (giving negative SN yields). 
For very massive stars 
($M \gtrsim 60 \,M_{\odot}$), rotating stars enter into the WR regime 
in the course of the
MS. In particular, the long time spent in the WNL phase
(WN star showing hydrogen at its surface \cite{ROTX})
 results in the ejection of large amounts of
H--burning products. Rotation therefore increases the H--burning
product yields in this mass range.

Concerning He--burning products, below 40 $M_{\odot}$, most of the 
$^{12}$C comes from the SN contribution.
In this mass range, rotating models, having larger cores, also have larger
yields (factor 1.5--2.5). 
For very massive stars 
($M \gtrsim 60 \,M_{\odot}$), 
the situation is reversed for He--burning products because of 
the different mass loss history.
As said above, rotating stars enter into the WR regime in the course of the
MS. The long time spent in the WNL phase
\cite{ROTX} results in a large mass loss. 
Therefore, very massive rotating stars have a small total mass early in
their evolution and end up with smaller cores.
Compared to 1992 \cite{AM92}, the $^{12}$C yields are larger in the 
present
rotating models for masses lower than 30 $M_{\odot}$ and 
smaller for masses higher than 30 $M_{\odot}$. 
Since very massive stars are much less numerous, we expect the
overall $^{12}$C yield of rotating models to be larger than those of
1992 \cite{AM92}. 
The situation for $^{16}$O and the total metallic yields is similar 
to carbon. Therefore $^{16}$O and metallic
yields are usually larger for our rotating models than for our 
non--rotating ones by a factor 1.5--2.5. 

\end{document}